\documentclass[submission,copyright,creativecommons]{eptcs}
\providecommand{\event}{FSFMA 2014} 
\usepackage{breakurl}             
\usepackage{epsfig}
\usepackage[sort]{cite}
\usepackage{algorithmic}
\usepackage{graphicx}
\usepackage[dvips,all]{xy}

\title{Experience using Coloured Petri Nets to Model Railway Interlocking Tables\thanks{Supported by National Research Council of Thailand Grant no. PorKor/2551-153.}}
\author{Somsak Vanit-Anunchai
\institute{
School of Telecommunication Engineering\\
Institute of Engineering\\
Suranaree University of Technology\\
Muang, Nakhon Ratchasima, Thailand}
\email{somsav@sut.ac.th}
}

\def\titlerunning{Experience using Coloured Petri Nets to Model Railway Interlocking Tables}
\def\authorrunning{S. Vanit-Anunchai}

\begin{document}
\maketitle

\begin{abstract}
Interlocking tables are the functional specification defining the routes on which the passage of the train is allowed. Associated with the route, the states and actions of all related signalling equipment are also specified. It is well-known that designing and verifying the interlocking tables are labour intensive, tedious and prone to errors.
To assist the verification process and detect errors rapidly, we formally model and analyse the interlocking tables using Coloured Petri Nets (CPNs).
Although a large interlocking table can be easily modelled, analysing the model is rather difficult due to the state explosion problem and undesired safe deadlocks. The safe deadlocks are when no train collides but the train traffic cannot proceed any further.
For ease of analysis we incorporate automatic route setting and automatic route cancelling functions into the model. These help reducing the number of the deadlocks. We also exploit the new features of CPN Tools; prioritized transitions; inhibitor arcs; and reset arcs. These help reducing the size of the state spaces. We also include
a fail safe specification called flank protection into the interlocking model.
\end{abstract}

\section{Introduction} \label{sect:intro}
\subsection{Background} In the railway signaling domain, an interlocking table is a tabular representation comprising the sections or routes that the train is allowed to enter together with the required states and actions of all related equipment along the routes.
The interlocking tables play such an important role that operating procedures and train movement must be complied with it.
This document also acts as a legal agreement between the railway administrators and the contractors.
Railway signalling contractors usually have software tools generating the interlocking table from the track layout and track side equipments. However the generated table is not unique. It depends upon the signalling principle or regulation of each railway administrator.
After the interlocking tables are designed and checked by the contractors, they need to be rechecked by signal engineers. In the past
we manually inspected the submitted interlocking tables without any software tools. Thus the checking process was very slow, labour intensive and prone to errors.
To reduce the manpower and time consumed in the checking process we have introduced the State Railway of Thailand to the formal methods and CPN Tools since 2009.
\subsection{Motivation}
Previously, we modelled and analysed in \cite{CPN09} a single track railway station
using Coloured Petri Nets (CPNs).
We created a static model where CPN structure was used to mimic the signalling layout and the train movements.
A generic CPN model of the signal operation was also developed.
The content of the interlocking table coded into ML functions which are used on arc inscriptions. Modelling interlocking tables of other railway stations was simply done by changing the content of the ML functions. These ML functions were automatically generated. After the contractors submitted the interlocking table files in Microsoft-EXCEL format,
the tables were transformed to XML and then to ML functions using Extensible Stylesheet Language Transformations (XSLT).
The interlocking tables were formally verified by exhaustively searching for the states where trains collide.
\cite{CPN09} had two important problems. Firstly, when we had many signalling devices working together, the CPN diagram became too complicated. It took 2-3 days to create a new CPN model of the signalling layout for a large station. Secondly, \cite{CPN09} focused on only interlocking tables. Although the system was safe, the signalman could give the sequences of instructions (route setting) that leaded the train traffic into deadlocks. Using state space generation, our CPN model generated a lot of safe terminal markings that had no train collision but the train traffic was in deadlock. It was very inconvenient to investigate all terminal markings in \cite{CPN09}.

To solve the first problem, we modeled in \cite{Coordinate2010} the signalling layout by encoding the geographic information into \emph{tokens} with a complex data structure.
When signaling layout was modified or rebuilt, we simply change the initial marking.
To solve the second problem, this paper introduces the automatic route setting and automatic route canceling functions into the CPN model. Although these two procedures are not specified in the interlocking tables, both are standard operating procedures normally conducted by signalmen.
After applying these two procedures, the sequences of route setting commands that lead to traffic deadlocks could be avoided.
\vspace{-4mm}
\subsection{Contributions}
The contribution of this paper is three fold.
Firstly, to ease of analysis and get rid of the undesired, safe terminal markings, the automatic route setting and automatic route canceling functions are included into the model.
Secondly, when we analysed the double track station in \cite{Coordinate2010}, we encountered the state explosion.
To alleviate the state explosion problem, we revise the CPN model by exploiting the recently introduced features of CPN Tools version 4.0.0 \cite{Westergaard13} that are prioritized transitions; inhibitor arcs; and reset arcs.
Thirdly,
designing a large interlocking table is a difficult task partly because the railway signalling system is required to be \emph{fail safe}.  The \emph{fail safe} means that, in the event of failure, the system shall respond in a no harmful way or no danger to persons. In the railway signalling domain an important event of failure is when a train overruns the stop signal.
Preventing an accident when a train overruns the stop signal, a fail safe condition called ``Flank Protection'' is required.
This condition has not been included in \cite{CPN09} and \cite{Coordinate2010} because it does not affect the normal functional behaviour as long as no fault occurs in the braking system and the train driver still obeys the signal. However this paper has included the flank protection into the model.

The rest of this paper is organised as follows. Section~2 briefly explains the concept of railway signalling system and interlocking tables. Section~3 reviews related work. The CPN model is discussed in Section~4. Analysis results are reported in Section 5. Section~6 presents conclusion and outlines suggested future work.

\vspace{-2mm}
\section{Railway Signalling Systems and Interlocking Tables}
\label{sect:RSS-Tables}
\subsection{Signalling Systems}
\label{sect:SS}
\begin{figure} [t!]
\centering
\includegraphics [width=15.8cm] {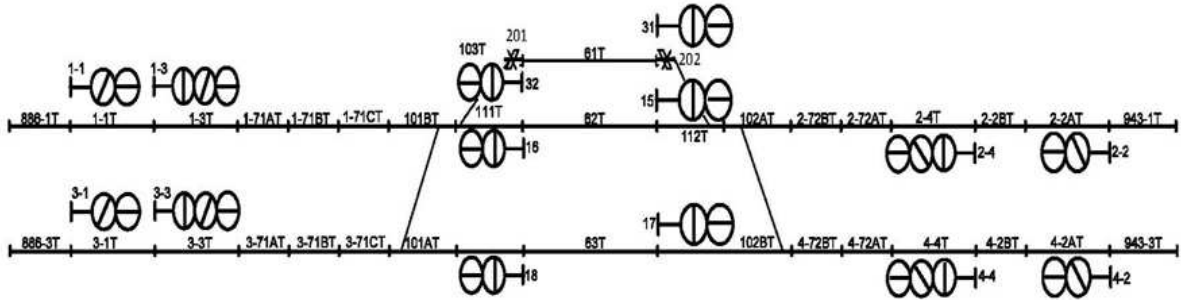}
\vspace{-5mm}
\caption{Signalling layout of the Panthong Station (double track)}
\label{fig:layout}
\vspace{-5mm}
\end{figure}

In general the railway lines are divided into \emph{sections}.
To avoid collision, only one train is allowed in one \emph{section} at a time. The train can enter or leave the \emph{section} when the driver receives authorization from a signalman via a signal indicator. Before the signalman issues the authorization, he needs to ensure that no object blocks the passage of the train.
The \emph{section} between two railway stations, which involves two signal men, is called ``\emph{block section}''.
To prevent human error that may lead to collisions, the strict operation on a \emph{block section} is controlled by equipment called ``Block Instruments''.
Figure~\ref{fig:layout} shows the signalling layout of a double track station named ``Panthong''.
The signalling layout comprises a collection of railway tracks and signalling equipment such as track circuits, points and signals.
(e.g.~signal no.1-3 and signal no.2-4).
Each piece of signalling equipment has an identification number and holds a certain state as follows.

\indent {\emph{\textbf{Track Circuits}}}
A track circuit is an electrical device used to detect the presence of a train. A track circuit (e.g.~61T, 1-3T) is either \emph{clear} indicating no train on the track or \emph{occupied} indicating the possible\footnote{When the track circuit fails, its state is occupied even if there is no train.} presence of a train.
\\
\indent{\emph{\textbf{Warner signals}}}
A warner signal (e.g.~1-1, 2-2, 3-1,4-2) has two aspects: \emph{yellow} or \emph{green}. It
informs drivers about the status of the next signal.
\\
\indent{\emph{\textbf{Home signals}}}
A home signal (e.g.~1-3, 2-4, 3-3, 4-4) has three aspects: \emph{red}, \emph{yellow} or \emph{green}.
It displays \emph{red} when the train is forbidden to enter the \emph{station area}.
It displays \emph{yellow} giving the driver authority to move the train into the \emph{station area} and prepare to stop at the next signal.
It displays \emph{green} giving the driver authority to move the train passing the \emph{station} and enter the next \emph{block section}.
\\
\indent{\emph{\textbf{Starter signals}}}
A starter (e.g.~15, 16, 17, 18, 31, 32) has two aspects: \emph{red} or \emph{green}.
It displays \emph{red} when forbidding the train to enter the \emph{block section}.
It displays \emph{green} when giving the driver authority to move the train into the \emph{block section}.
\\
\indent{\emph{\textbf{Point}}}
A point (e.g.~101A, 101B, 111, 112, 102A, 102B) or railway switch or turnout is a mechanical installation used to guide a train from one track to another.
A point usually has a straight through track called ``main-line'' and a diverging track called loop line.
A point is right-hand when a moving train from a joint track diverges to the right of the straight track.
Similarly a left-hand point has the diverging track on the opposite side of a right-hand point.
When a point diverges the train, it is in reverse position. When a point lets the train move straight through, it is in normal position.

\indent{\emph{\textbf{Derailer}}}
A derailer (e.g.~201, 202)
is a mechanical installation used to prevent unauthorized movements of trains or unattended rolling stock. The train is derailed when it rolls over the derailer. The normal position is the derailing position.

\subsection{Interlocking Tables}
\label{CT}
A collection of track circuits along the reserved \emph{section} is called ``\emph{route}''.
An entry signal shall be clear to let the train enter the route.
Although the request to clear the entry signal is issued by the signalman, the route entry permission is decided by the interlocking system
using safety rules and control methods specified in the agreed Interlocking Tables.
Tables~\ref{fig:C_Table_1},~\ref{fig:C_Table_2},~and~\ref{fig:C_Table_3}   are the Interlocking Tables (partial) for Panthong station of which the signalling layout is shown in Fig.~\ref{fig:layout}.
Data in the first column, ``From'', is the route identifications which are labelled by the entry signal: 1-3(1); 1-3(2);  3-3(1); 3-3(2); 3-3(3);
              2-4(1); 2-4(2);  4-4(1); 4-4(2); 4-4(3);
              15(1); 15(2); 16(1); 16(2);  31(1);31(2); 32(1);32(2); 17 and 18. Due to space limitation we show only 4 routes in Tables \ref{fig:C_Table_1}, \ref{fig:C_Table_2} and \ref{fig:C_Table_3}.
Each row in the tables represents the requirement how to set and release each route.
For example, route 3-3(3) comprises the track circuits 3-3T, 3-71AT, 3-71BT,3-71CT,101AT, 18T, 63T, 17T and requires that the
point 101 is in normal position.
Routes 3-3(1), 3-3(2) and 3-3(3) distinguish that behind signal 3-3
three routes are possible. Similar rule applies to routes 1-3; 2-4; and 4-4.
The column ``Requires Route Normal'' shows conflict routes. A route cannot be set
if any conflicting routes have been set and not yet released.
For route 3-3(3) the conflicting routes are 16(2), 32(2), 3-3(1), 3-3(2), 18, and 4-4(3).
The exit (starter) signal of this route is 17,
and if home signal 3-3 shows green, then starter signal 17 shows green.

Different Interlocking systems from different manufacturers may have different control methods.
However there are four basic control methods widely accepted and used among railway companies.

\begin{table} [b!]
\begin{center}
\caption{\normalsize An Interlocking Table for Panthong station (part 1:Route locking)}
\label{fig:C_Table_1}
\includegraphics[width=15cm]{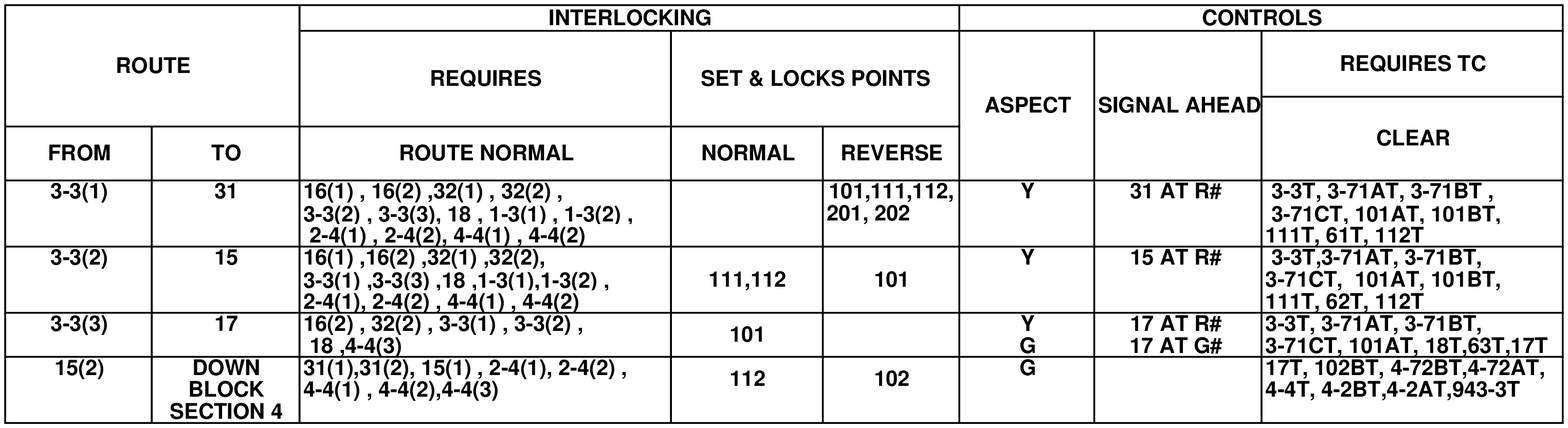}
\end{center}
\end{table}

\begin{table} [t]
\begin{center}
\caption{\normalsize An Interlocking Table for Panthong station (part 2:Approach locking)}
\label{fig:C_Table_2}
\includegraphics[width=15cm]{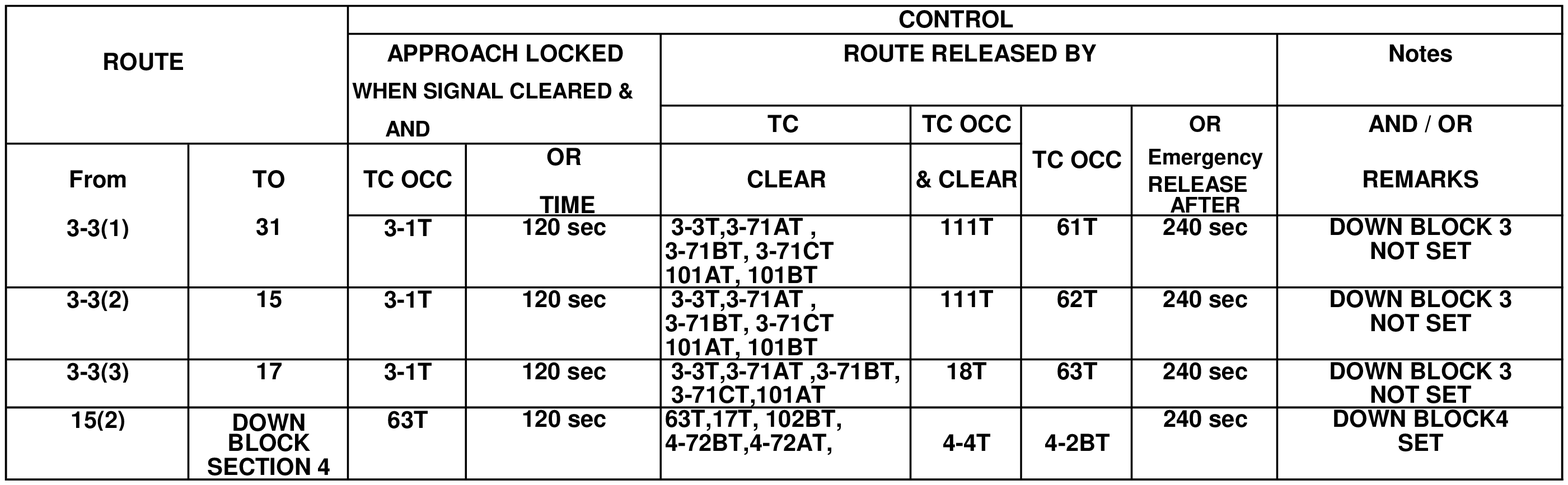}
\vspace{-5mm}
\end{center}
\end{table}

\vspace{-0mm}

\begin{table} [t]
\begin{center}
\caption{\normalsize An Interlocking Table for Panthong station (part 3:Flank Protection)}
\label{fig:C_Table_3}
\includegraphics[width=10cm]{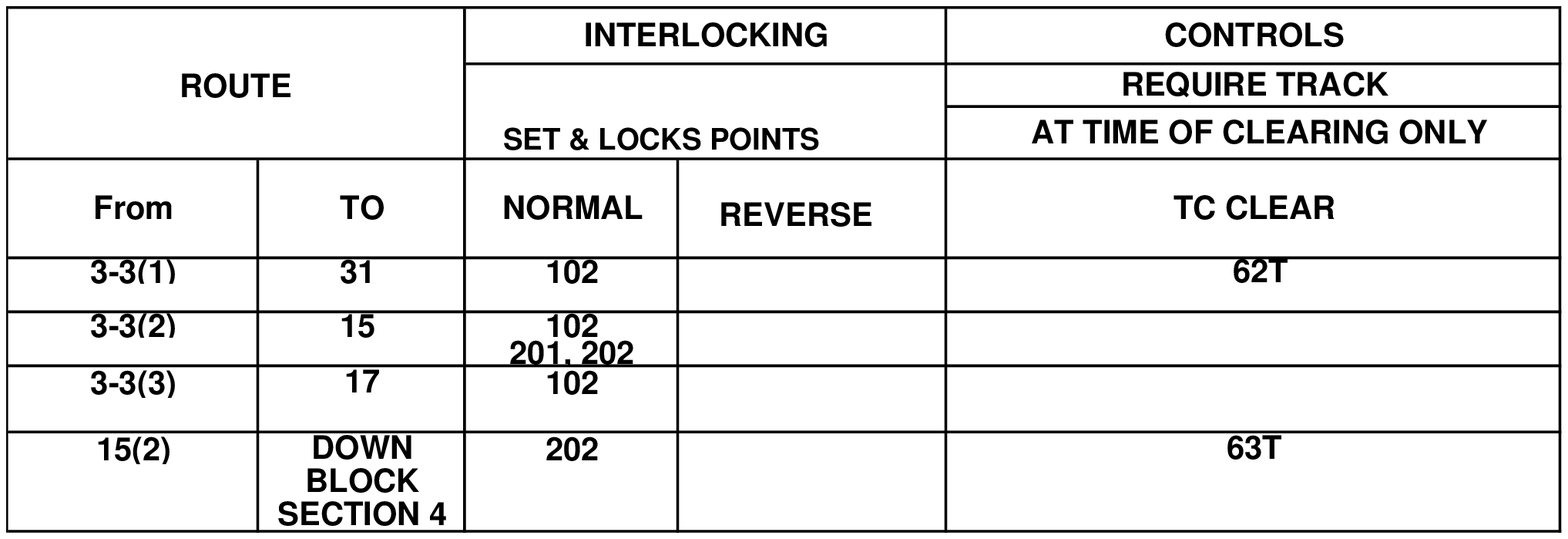}
\vspace{-5mm}
\end{center}
\end{table}

\indent  {\textbf{Route locking}}
Route setting involves a collection of adjacent track circuits, points and signals.
A route can be set and reserved for a passage of a train along this route.
To assure the safety, firstly, the interlocking system verifies that the route does not
conflict with other routes previously set. Secondly, the points along the route are locked in the correct positions. If the related points are not in the correct positions, the controller will attempt to set and lock them in the correct positions. Thirdly, the track circuits along the required route are all clear or unoccupied so that nothing obstructs the passage of the train.  Then the entry signal can be cleared (showing yellow or green).

\indent {\textbf{Approach locking}} After a route is set; the point is locked; and the entry signal is
cleared, if the track circuit in front of (approaching) the entry signal is occupied,
then the signalman cannot cancel the route and the entry signal by the normal procedure.  Approach locking prevents the train driver from the sudden change of signal aspect from green or yellow to red. Column 3 in Table~\ref{fig:C_Table_2}, ``APPROACH LOCKED WHEN SIGNAL CLEARED \& TC OCC'',  presents locking when a route is set and the approach track circuit is occupied. For example, route 3-3(3) will be approach locked if the route is set and track 3-1T is occupied.

\indent{\textbf{Route released}} After the passage of the train, the reserved route is released automatically. Column ``Route Released by'' in Table \ref{fig:C_Table_2}  presents route released  mechanism for the signalling layout in Fig.~\ref{fig:layout}. Route 3-3(3) will be released when the track circuits 3-3T, 3-71AT, 3-71BT, 3-71CT, 101AT are clear; the track circuit 18T is occupied and then clear; and the track circuit 63T is occupied. The reserved route can be emergency released but the release action will be delayed for 4 minutes after the signalman issues the ``emergency route released'' command.

\indent{\textbf{Flank protection}} This is an important class of fail safe requirement. The equipment within the surrounding area of the reserved route that may cause an accident shall be protected even if no train is expected to pass such a signal or such points. Points should be in such positions that they do not give immediate access to the route. Even though those flank points and derailers are not located on the required route, when the route is set, they shall be locked in the safe position until the route is released. Table~\ref{fig:C_Table_3} shows the flank protection requirements for routes 3-3(1), 3-3(2), 3-3(3) and 15(2) of the Panthong's interlocking table.
For example route 3-3(3) requires the points 102 (both 102A and 102B), which are not on the route, be locked in the normal position.
Route 15(2) requires the track circuits 63T, which is not in the route 15(2), be unoccupied.
Route 15(2) also requires the derailer 202, which is not in the route 15(2), be locked in the derail position.
Because routes 3-3(3) and 15(2) are not in conflict, trains may enter these two routes at the same time. However arriving on 63T, the train on route 3-3(3)
could overrun the red signal no.~17 and collide with the train on route 15(2) at point 102.
To prevent this accident, route 3-3(3) requires flank protection, point 102 be locked
in the normal position. Meanwhile Route 15(2) cannot be set while point 102 in the normal position.

\section{Related Work}
\label{RelatedWork}
In \cite{WJFPRH:1998}, Fokkink and Hollingshead divide the railway signalling system into three layers: infrastructure, interlocking and logistics layers. The infrastructure layer involves objects or equipment used in the yard. The work in this category, for instance \cite{CCDCPS:2008,Czarnecki}, ties closely with the manufacturer's products.
The logistics layer involves human operation and train scheduling which aims at efficiency and absence of deadlocks. It involves the operation of the whole railway network (e.g.\cite{AMH:2007,Jan:1998}) thus the state space explosion problem is often encountered.
The interlocking layer provides the interface between logistics and infrastructure layers. It prevents us from accidents caused by human errors or equipment failure.  The work in this category models the interlocking tables and verifies them against the signalling principles. For example \cite{WJFPRH:1998,WKWWJ2005:ACSC} uses theorem prover and \cite{KWNR2003:ACSC} uses NuSMV.
Hansen \cite{Hansen:1994a} presented a VDM model of a railway interlocking
system, and validate it through simulation using Meta Language (ML). The work focuses on the principles and concepts of Danish systems rather than a particular interlocking system. He also pointed out that Interlocking systems from other countries may be different from the Interlocking described in \cite{Hansen:1994a}.
Winter et al \cite{WKWWJ2002:ACSC} proposed to create two formal models during  the design process of interlockings.
One is the formal model of the Signalling Principles called Principle model.  The other is the formal model
of the functional specification for a specific track-layout called Interlocking model.
The Control Tables are  translated  into an interlocking model and  then checked against the Principle model.
At first she used CSP (Communicating Sequential Processes) as a modelling language but later found that
the CSP models of the interlocking system and the signalling principle are difficult
to understand and validate. Thus \cite{KWNR2003:ACSC} used ASM (Abstract State Machine) notation to model the semantics of control tables. The ASM model is then automatically transformed to NuSMV code \cite{NuSMV} while the safety properties are modeled in CTL (Computational Tree Logic).
Basten \cite{basten95} simulated and analysed railway interlocking specification using ExSpect which is  a software tool based on high level Petri Nets. However formal verification of railway interlockings were not possible because they were too complex for the technology at that time.
Hagalisletto et al \cite{AMH:2007} modelled signalling equipment such as track circuits and turnouts using Coloured Petri Nets. But their aim is to simulate the train schedule rather than to verify the interlocking.

\section{The CPN Model of the Panthong's Interlocking Table}
\label{sect:model}
Coloured Petri Nets (CPNs)~\cite{KJ:CPNsttt2} are a graphical modelling language for design, verification and analysis of distributed, concurrent and complex systems. CPNs include hierarchical constructs that allow modular specifications to be created. CPN Tools~\cite{KJ:CPNsttt2} is a software tool used to create, maintain, simulate and analyse CPNs. We use CPN Tools version 4 \cite{Westergaard13} to create our railway signaling model and analyse them using reachabilty analysis. 

\subsection{Modelling Scope and  Assumptions}
To reduce the complexity of the model as well as avoid the state explosion problem when analysing railway networks \cite{KWNR2003:ACSC,AMH:2007},
we need to make the following assumptions regarding train movement and signalling operations:

\begin{enumerate}

\item We assume that a train has no length and it occupies one track at a time. The train moves in only one direction. Train shunting is not considered.
\item Our model does not include the auxiliary signals such as Call-on, Shunting and Junction indicators.
\item Our model does not include level crossings.
\item Our model includes high level abstraction of block systems but we do not model their operations in detail.
\item Our model does not include timers.
\item The train must not move through a track circuit so fast that the interlocking cannot detect the presence of the train. We use \emph{prioritized transitions} to model this condition.
\item Unlike \cite{Coordinate2010}, our CPN model includes the flank protection.
\end{enumerate}

\subsection{Examples of the CPN Model}
This section provides two examples of CPN pages. Due to space limitation we choose to explain only the \texttt{UserCommand} and \texttt{Move\_Track\_ to\_Track} pages because these pages play central roles in the model. For global declarations
and other details regarding our CPN model of the interlocking
table, see \cite{CPN09} and \cite{Coordinate2010}.

\subsubsection{UserCommand page}
\begin{figure} [b!]
\centering
\includegraphics[width=12cm]{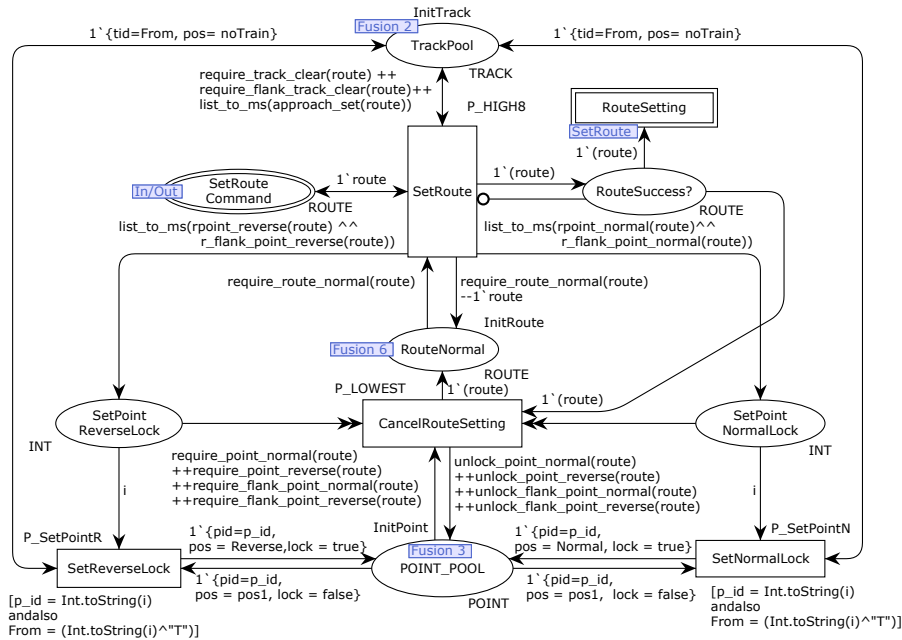}
\vspace{-2mm}
\caption{CPN model: UserCommand page}
\label{fig:UserCommand}
\vspace{-3mm}
\end{figure}

The \texttt{UserCommand} page shown in Fig.~\ref{fig:UserCommand} models the action after a route request command is issued (e.g. 3-3(3)).  Transition \texttt{SetRoute} checks whether it is plausible to set the requested route.
Taking tokens from fusion places \texttt{RouteNormal} and \texttt{TrackPool}, transition \texttt{SetRoute} checks if

\begin{enumerate}

\item No conflict route is being set (modelled by function require\_route\_normal).
\item The relevant tracks are unoccupied (modelled by functions require\_track\_clear and \linebreak require\_flank\_track\_clear).

\end{enumerate}

If all conditions are met, transitions \texttt{SetNormalLock} and \texttt{SetReverseLock} will attempt to set and lock points in the correct position. The two conditions and the states of the relevant point machines and derailers will be rechecked again by Substitution transition \texttt{RouteSetting}.

Actually the above model description is enough to satisfy the specification requirement.
However when the CPN model was analysed, we found many deadlocks which were safe terminal states.
It is inconvenient to investigate all deadlocks so we attempt to reduce them by
introducing automatic route setting and automatic route canceling. These two functions are not specified in the interlocking table because they are normally conducted by the signal man.
The automatic route setting condition is that the preselected route setting command can be issued only when the track in front of the entry signal is occupied. This is modelled by the ML function \texttt{approach\_set}.
After the transition \texttt{SetRoute} fires,
it will be disabled by the inhibitor arc from place \texttt{RouteSuccess?}.
When the route setting process is not complete, no other route can be set.
Transitions \texttt{SetNormalLock} and \texttt{SetReverseLock} attempt to set and lock the points in the position specified in the interlocking table. Because transition \texttt{SetNormalLock} has a higher priority, the actions of transition \texttt{SetNormalLock} do not interleave with the actions of transition \texttt{SetReverseLock}.

When the route setting cannot be completed and no more action can occur in the model,
the incomplete route setting will be canceled using transition \texttt{CancelRouteSetting}
(lowest priority). This transition clears all tokens in the places \texttt{SetPointReverseLock} and \texttt{SetPointNormalLock} by the reset arcs (two arrow arcs) in a single instance.
Using prioritized transitions, inhibitor arcs and reset arcs can alleviate the state explosion problem.
The automatic route setting and automatic route cancelation can eliminate deadlocks due to the wrong sequence of route setting commands given by the signal man.

\begin{figure} [b!]
\centering
\includegraphics[width=11cm]{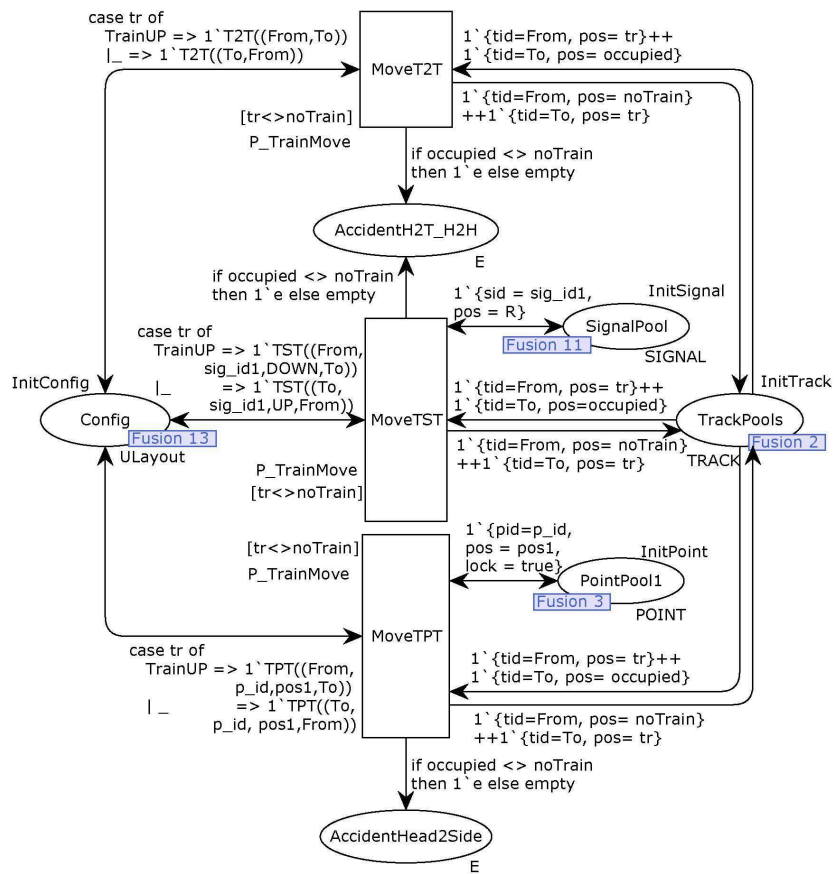}
\vspace{-2mm}
\caption{CPN model: Move Track to Track page}
\label{fig:MoveT2T}
\vspace{-3mm}
\end{figure}

\subsubsection{Move\_Track\_ to\_Track page}
Figure~\ref{fig:MoveT2T} shows the CPN diagram modelling the simple train movements between two adjacent tracks.
Place \texttt{Config} stores tokens representing signalling layout as discussed in \cite{Coordinate2010}. In addition to the layout, the train movement requires information regarding the status of signalling equipment stored in places \texttt{TrackPools}, \texttt{SignalPool}, and \texttt{PointPool1}.
Transition \texttt{MoveT2T} represents the movement across adjacent straight tracks. Transition \texttt{MoveTST} behaves similar to Transition \texttt{MoveT2T} but there is an entry signal post between the adjacent tracks. However the train moves toward the back of the signal.
The train movement facing the front of the entry signal was modelled in another CPN page illustrated in \cite{Coordinate2010}. The movement across points is captured by Transition \texttt{MoveTPT}. For ease of analysis we also add two  places \texttt{AccidentH2T\_H2H} and \texttt{AccidentHead2Side} for detecting train collision.

\section{Analysis}
\subsection{Desired Property}
A basic safety property that railway signalling shall provide is to prevent train collision and derailment. Places \texttt{AccidentH2T\_H2H} and \texttt{AccidentHead2Side} shall be empty when no collision occurs. Checking derailment is in other CPN pages that we do not discussed in this paper.
To convince us of the correctness of our CPN model and the interlocking table, the CPN model is analysed using reachability analysis in CPN Tools version 4.0.0. The investigation of the generated reachability graph is conducted on Windows XP using a AMD9650 computer with 2.30 GHz and 3.5 GB of RAM. After generating each entire graph, we use ML query functions searching
for the markings that have tokens in places\footnote{Of course we also need to check other accident places in other CPN pages that are not discussed in this paper.}
\texttt{AccidentH2T\_H2H} or \texttt{AccidentHead2Side}.
For ease of investigating the terminal markings, we attempt to execute the model until there is no train in the model. This can be done using automatic route setting and automatic route cancelation.
However there are still possible deadlocks left as shown in Section 5.3.

\subsection{Initial Configurations}
Despite the fact that we can analyse various scenarios by changing the
initial markings, due to space limitation, we select to discuss only six cases with the initial configurations shown in Table~\ref{tab:INITMARK}.
The initial configurations are:
\begin{enumerate}
\item Case A is when three trains are on the platform tracks.
\item Case B is when two trains are on the platform tracks 62T and 63T.
\item Case C1,C2, and C3 are when one train is on the platform track 61T, 62T, and 63T respectively.
\item Case D is when no train is on the platform tracks.
\end{enumerate}

In all initial markings, four trains are coming from the north and south directions and other track circuits are unoccupied; all points are in \texttt{Normal} position and unlocked; all derailers are \texttt{Normal} and locked.  All signals are in normal states.

\begin{table} [b!]
\begin{center}
\caption{\normalsize Initial configurations of track circuits.}
\label{tab:INITMARK}
\includegraphics[width=12cm]{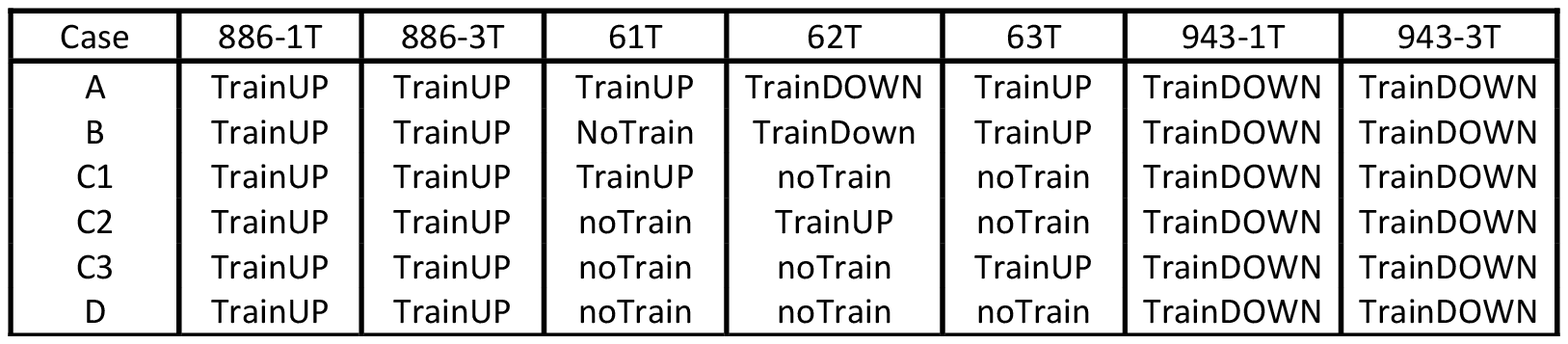}
\vspace{-5mm}
\end{center}
\end{table}

\subsection{Analysis Results}
Tables \ref{tab:cmp_size} and \ref{tab:size} show analysis results: state space sizes; execution time; and the number of deadlocks. All markings are safe (no train collision).
In particular, Tables \ref{tab:cmp_size} illustrates that our approach can reduce the state space sizes.

\begin{enumerate}
\item B[Coor2010] was the old analysis result of Case B (from \cite{Coordinate2010}).
\item B[no Flank Protection] is a new result of Case B when the CPN model is revised not only using prioritized transitions; inhibitor arcs; reset arcs but also including the automatic route setting and automatic route canceling functions. However the model has not included the flank protection.
This result shows that our proposed reduces the number of states to about 70\%. The number of terminal markings is also reduced significantly.
\item B[with Flank Protect] is the result of Case B when we add the flank protection into the model. Because of this restriction, the non-conflicting routes in \cite{Coordinate2010} that has no overlapped section now become conflicted so that the state space size is reduced drastically .
\end{enumerate}

Revising model structure with the flank protection requirement, we are able to analyse the scenarios that we cannot reach before (Case C1, C2, C3 and D).
The details of the terminal markings are listed in Table \ref{tab:result3}. They show the occupancies of trains on the tracks in front of the entry signal. In all terminal markings other tracks are unoccupied.  All signals are in the normal states.
Terminal markings no. 5 of Case C1 and no. 7 of Case D suggest that the signal man can manage the traffic such that no deadlock occurs. For the traffic of Case C2 and C3 there always be deadlocks so that the emergency procedure shall be carefully conducted to solve the deadlocks.

\begin{table} [h]
\begin{center}
\caption{\normalsize Comparison of the state space sizes (with \cite{Coordinate2010}).}
\label{tab:cmp_size}
\includegraphics[width=10cm]{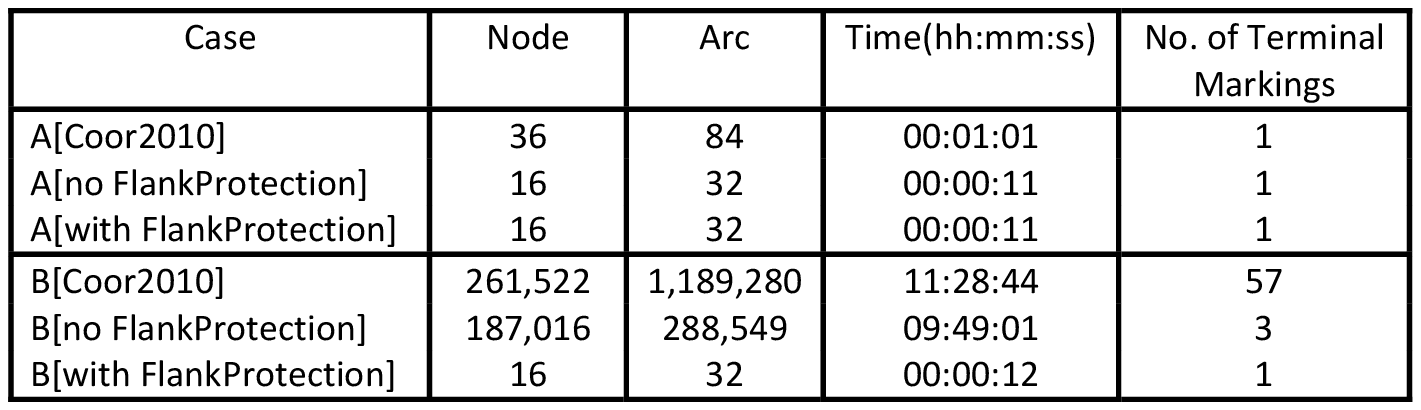}
\vspace{-5mm}
\end{center}
\end{table}

\begin{table} [h]
\begin{center}
\caption{\normalsize Summary of state space results.}
\label{tab:size}
\includegraphics[width=10cm]{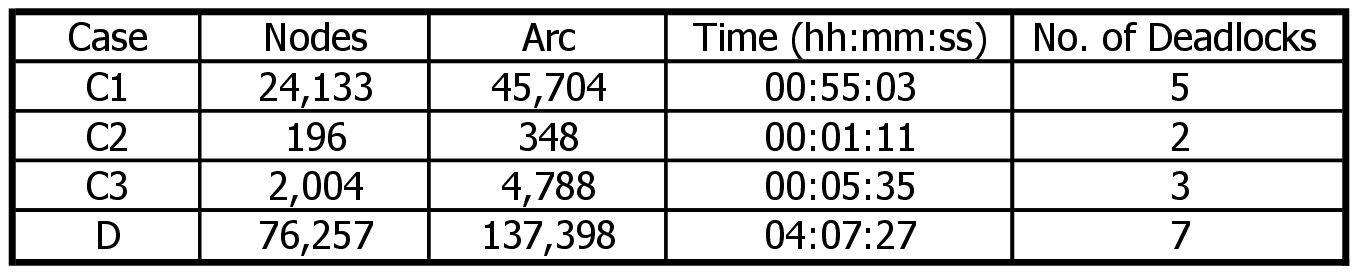}
\vspace{-7mm}
\end{center}
\end{table}

\begin{table} [h]
\begin{center}
\caption{\normalsize Terminal Markings.}
\label{tab:result3}
\includegraphics[width=13cm]{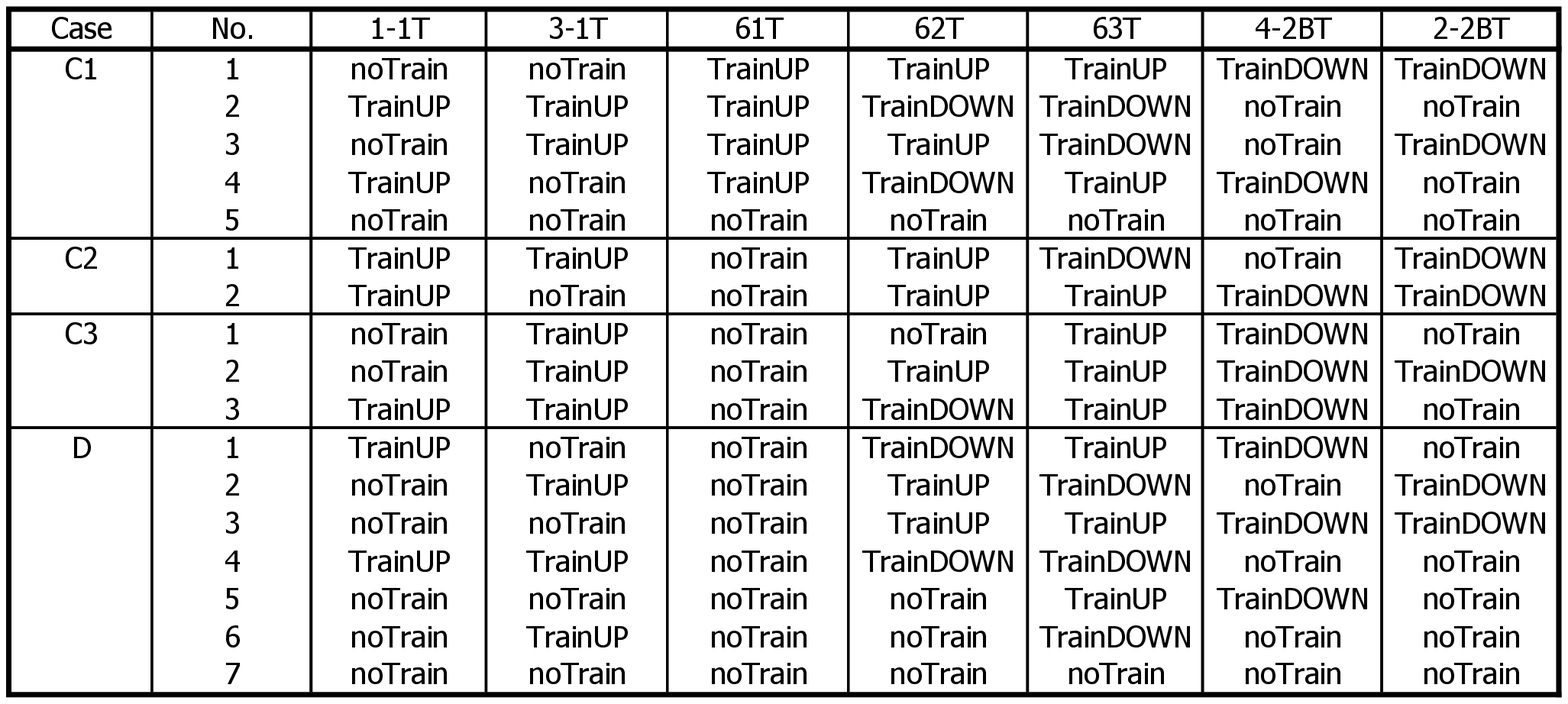}
\vspace{-5mm}
\end{center}
\end{table}

\section{Conclusion and Suggested Work}
This paper restructures the previous CPN model in \cite{Coordinate2010} to make the analysis process easier and alleviate the state explosion problem. Our study shows that adding the automatic route setting and automatic
route canceling functions into the CPN model reduces the number of undesired deadlocks.
These two functions are not specified in the interlocking table but they are normally conducted by the signalmen.
Our study also shows that using prioritized transitions, inhibitor arcs and reset arcs
can reduce the state space sizes.
Although the flank protection significantly reduces the state space size, we discover that it masks out errors in the route locking part of the interlocking table (Table \ref{fig:C_Table_1}). It seems inevitable that the verification of the interlocking table without the flank protection shall be conducted before the flank protection functions are verified.

From a modelling perspective it is easy to add the flank protection requirement but from analysis perspective it is not so easy to be verified. The flank protection is a fail safe requirement preventing an accident when equipment fails or a train passes a signal at danger. This dangerous scenario normally cannot be reached in our regular CPN model. To verify the flank protection and reach the states that are normally unreachable, the CPN model needs to allow the train pass a signal at danger. Thus, we suggest to conduct experiments by deleting the signal from the signalling layout and let the train pass over. This can be easily done by modify the configuration tokens that represent geographic information. This \emph{modified} CPN models are used to generate the reachability graphs. When we search the entire graphs, we expect no train collision.
However, if those points are not related to the required route at all, accidents definitely do not occur regardless of the point positions either normal or reverse. Thus, to prove the safety properties of the flank protection requirement we need to prove two properties. Firstly, if the flank protection works correctly, no train collision occurs. Secondly, if the CPN model does not include the flank protection, trains will collide.

When verifying the flank protection in the interlocking table, we always assume that the flank points are known. However, for a large and complex station layout, it is difficult to identify the flank points without any errors. To facilitate the design and verification tasks, we suggest to use the modified CPN model generating train collision scenarios. Tracing the markings before trains collide should help us identify the flank points and their correct positions.

\paragraph {Acknowledgments}
The author is thankful to the anonymous reviewers. Their constructive feedback has helped the author improve the quality of this paper.


\nocite{*}
\bibliographystyle{eptcs}

\bibliography{ref-FSFMA-2014}

\begin{thebibliography}{10}
\providecommand{\bibitemdeclare}[2]{}
\providecommand{\surnamestart}{}
\providecommand{\surnameend}{}
\providecommand{\urlprefix}{Available at }
\providecommand{\url}[1]{\texttt{#1}}
\providecommand{\href}[2]{\texttt{#2}}
\providecommand{\urlalt}[2]{\href{#1}{#2}}
\providecommand{\doi}[1]{doi:\urlalt{http://dx.doi.org/#1}{#1}}
\providecommand{\bibinfo}[2]{#2}

\bibitemdeclare{inproceedings}{Czarnecki}
\bibitem{Czarnecki}
\bibinfo{author}{A.~Svendsen \surnamestart et~al\surnameend}
  (\bibinfo{year}{2008}): \emph{\bibinfo{title}{The Future of Train
  Signaling}}.
\newblock In: {\sl \bibinfo{booktitle}{Proceedings of MoDELS 2008}}, {\sl
  \bibinfo{series}{Lecture Notes in Computer Science}} \bibinfo{volume}{5301},
  \bibinfo{publisher}{Springer Verlag}, pp. \bibinfo{pages}{128--142},
  \doi{10.1007/978-3-540-87875-9\_9}.

\bibitemdeclare{article}{basten95}
\bibitem{basten95}
\bibinfo{author}{T.~\surnamestart Basten\surnameend},
  \bibinfo{author}{R.~\surnamestart Bol\surnameend} \&
  \bibinfo{author}{M.~\surnamestart Voorhoeve\surnameend}
  (\bibinfo{year}{1995}): \emph{\bibinfo{title}{Simulating and Analyzing
  Railway Interlockings in ExSpec}}.
\newblock {\sl \bibinfo{journal}{IEEE Parallel \& Distributed Technology,
  Systems \& Applications}} \bibinfo{volume}{3}(\bibinfo{number}{3}), pp.
  \bibinfo{pages}{50--62}, \doi{10.1109/M-PDT.1995.414843}.

\bibitemdeclare{inproceedings}{MOCA2006}
\bibitem{MOCA2006}
\bibinfo{author}{J.~\surnamestart Bj{\o}rk\surnameend}, \bibinfo{author}{A.~M.
  \surnamestart Hagalisletto\surnameend} \& \bibinfo{author}{P.~\surnamestart
  Enger\surnameend} (\bibinfo{year}{June 2006}): \emph{\bibinfo{title}{Large
  Scale simulations of Railroad Nets}}.
\newblock In: {\sl \bibinfo{booktitle}{Proceedings of the Fourth International
  Workshop on Modelling of Objects, Components and Agents, MOCA'06,Bericht 272,
  FBI-HH-B-272/06}}, pp. \bibinfo{pages}{45--101}.

\bibitemdeclare{inproceedings}{CCDCPS:2008}
\bibitem{CCDCPS:2008}
\bibinfo{author}{C.~\surnamestart Chevillat\surnameend},
  \bibinfo{author}{D.~\surnamestart Carrington\surnameend},
  \bibinfo{author}{P.~\surnamestart Strooper\surnameend},
  \bibinfo{author}{J.~G. \surnamestart S{\"u}{\ss}\surnameend} \&
  \bibinfo{author}{L.~\surnamestart Wildman\surnameend} (\bibinfo{year}{2008}):
  \emph{\bibinfo{title}{Model-Based Generation of Interlocking Controller
  Software from Control Tables}}.
\newblock In: {\sl \bibinfo{booktitle}{Proceeding of ECMDA-FA 2008}}, {\sl
  \bibinfo{series}{Lecture Notes in Computer Science}} \bibinfo{volume}{5095},
  \bibinfo{publisher}{Springer, Heidelberg}, pp. \bibinfo{pages}{349--360},
  \doi{10.1007/978-3-540-69100-6\_24}.

\bibitemdeclare{inproceedings}{NuSMV}
\bibitem{NuSMV}
\bibinfo{author}{A.~\surnamestart Cimatti\surnameend},
  \bibinfo{author}{F.~Giunchiglia \surnamestart E.~Clarke\surnameend} \&
  \bibinfo{author}{M.~\surnamestart Roveri\surnameend} (\bibinfo{year}{1999}):
  \emph{\bibinfo{title}{NuSMV: A new symbolic model verifier}}.
\newblock In: {\sl \bibinfo{booktitle}{Proceedings of International Conference
  on Computer Aided Verification, CAV'99}}, {\sl \bibinfo{series}{Lecture Notes
  in Computer Science}} \bibinfo{volume}{1633}, \bibinfo{publisher}{Springer
  Verlag}, pp. \bibinfo{pages}{495--499}, \doi{10.1007/3-540-48683-6\_44}.

\bibitemdeclare{inproceedings}{WJFPRH:1998}
\bibitem{WJFPRH:1998}
\bibinfo{author}{W.J. \surnamestart Fokkink\surnameend} \&
  \bibinfo{author}{P.R. \surnamestart Hollingshead\surnameend}
  (\bibinfo{year}{May 1998}): \emph{\bibinfo{title}{Verification of
  Interlockings: from Control Tables to Ladder Logic Diagrams}}.
\newblock In: {\sl \bibinfo{booktitle}{Proceedings of 3rd Workshop on Formal
  Methods for Industrial Critical Systems (FMICS'98)}},
  \bibinfo{publisher}{Stichting Mathematisch Centrum},
  \bibinfo{address}{Amsterdam}, pp. \bibinfo{pages}{171--185}.

\bibitemdeclare{article}{AMH:2007}
\bibitem{AMH:2007}
\bibinfo{author}{A.~M. \surnamestart Hagalisletto\surnameend},
  \bibinfo{author}{J.~\surnamestart Bj{\o}rk\surnameend},
  \bibinfo{author}{I.~C. \surnamestart Yu\surnameend} \&
  \bibinfo{author}{P.~\surnamestart Enger\surnameend} (\bibinfo{year}{2007}):
  \emph{\bibinfo{title}{Constructing and Refining Large-Scale Railway Models
  Represented by Petri Nets}}.
\newblock {\sl \bibinfo{journal}{IEEE Transactions on Systems, Man, and
  Cybernetics, Part C}} \bibinfo{volume}{37}(\bibinfo{number}{4}), pp.
  \bibinfo{pages}{444--460}, \doi{10.1109/TSMCC.2007.897323}.

\bibitemdeclare{inproceedings}{Hansen:1994a}
\bibitem{Hansen:1994a}
\bibinfo{author}{K.~M. \surnamestart Hansen\surnameend} (\bibinfo{year}{1994}):
  \emph{\bibinfo{title}{Formalizing Railway Interlocking Systems}}.
\newblock In: {\sl \bibinfo{booktitle}{Nordic Seminar on Dependable Computing
  Systems}}, \bibinfo{publisher}{Department of Computer Science, Technical
  University of Denmark}, pp. \bibinfo{pages}{83--94}.

\bibitemdeclare{phdthesis}{Jan:1998}
\bibitem{Jan:1998}
\bibinfo{author}{C.~W. \surnamestart Janczura\surnameend}
  (\bibinfo{year}{1998}): \emph{\bibinfo{title}{Modelling and Analysis of
  Railway Network Control Logic using Coloured Petri Nets}}.
\newblock Ph.D. thesis, \bibinfo{school}{School of Mathematics and Institute
  for Telecommunications Research, University of South Australia},
  \bibinfo{address}{Adelaide, Australia}.

\bibitemdeclare{book}{KJ:CPNsttt2}
\bibitem{KJ:CPNsttt2}
\bibinfo{author}{K.~\surnamestart Jensen\surnameend} \& \bibinfo{author}{L.M.
  \surnamestart Kristensen\surnameend} (\bibinfo{year}{2009}):
  \emph{\bibinfo{title}{Coloured Petri Nets: Modelling and Validation of
  Concurrent Systems}}.
\newblock \bibinfo{publisher}{Springer, Heidelberg}, \doi{10.1007/b95112}.

\bibitemdeclare{inproceedings}{CPN09}
\bibitem{CPN09}
\bibinfo{author}{S.~\surnamestart Vanit-Anunchai\surnameend}
  (\bibinfo{year}{2009}): \emph{\bibinfo{title}{Verification of Railway
  Interlocking Tables using Coloured Petri Nets}}.
\newblock In: {\sl \bibinfo{booktitle}{the tenth Workshop and Tutorial on
  Practical Use of Coloured Petri Nets and the CPN Tools}},
  \bibinfo{series}{DAIMI PB 590}, \bibinfo{publisher}{Department of Computer
  Science, University of Aarhus}, pp. \bibinfo{pages}{139--158}.

\bibitemdeclare{inproceedings}{Coordinate2010}
\bibitem{Coordinate2010}
\bibinfo{author}{S.~\surnamestart Vanit-Anunchai\surnameend}
  (\bibinfo{year}{2010}): \emph{\bibinfo{title}{Modelling Railway Interlocking
  Table Using Coloured Petri Nets}}.
\newblock In \bibinfo{editor}{D.~\surnamestart Clarke\surnameend} \&
  \bibinfo{editor}{G.~\surnamestart Agha\surnameend}, editors: {\sl
  \bibinfo{booktitle}{Proceedings of the 12th International Conference on
  Coordination Models and Languages, (Coordination 2010)}}, {\sl
  \bibinfo{series}{Lecture Notes in Computer Science}} \bibinfo{volume}{6116},
  \bibinfo{publisher}{Springer, Heidelberg}, \bibinfo{address}{Amsterdam,
  Netherlands}, pp. \bibinfo{pages}{137--151},
  \doi{10.1007/978-3-642-13414-2\_10}.

\bibitemdeclare{inproceedings}{Westergaard13}
\bibitem{Westergaard13}
\bibinfo{author}{M.~\surnamestart Westergaard\surnameend}
  (\bibinfo{year}{2013}): \emph{\bibinfo{title}{{CPN Tools 4}: Multi-formalism
  and Extensibility}}.
\newblock In \bibinfo{editor}{Jos{\'e}~Manuel \surnamestart Colom\surnameend}
  \& \bibinfo{editor}{J{\"o}rg \surnamestart Desel\surnameend}, editors: {\sl
  \bibinfo{booktitle}{Petri Nets}}, {\sl \bibinfo{series}{Lecture Notes in
  Computer Science}} \bibinfo{volume}{7927}, \bibinfo{publisher}{Springer}, pp.
  \bibinfo{pages}{400--409}, \doi{10.1007/978-3-642-38697-8_22}.

\bibitemdeclare{inproceedings}{WKWWJ2002:ACSC}
\bibitem{WKWWJ2002:ACSC}
\bibinfo{author}{K.~\surnamestart Winter\surnameend} (\bibinfo{year}{2002}):
  \emph{\bibinfo{title}{Model Checking Railway Interlocking Systems}}.
\newblock In: {\sl \bibinfo{booktitle}{Proceeding of the 25th Australian
  Computer Science Conference (ACSC 2002)}}.

\bibitemdeclare{inproceedings}{WKWWJ2005:ACSC}
\bibitem{WKWWJ2005:ACSC}
\bibinfo{author}{K.~\surnamestart Winter\surnameend},
  \bibinfo{author}{W.~\surnamestart Johnston\surnameend},
  \bibinfo{author}{P.~\surnamestart Robinson\surnameend},
  \bibinfo{author}{P.~\surnamestart Strooper\surnameend} \&
  \bibinfo{author}{L.~\surnamestart van~den Berg\surnameend}
  (\bibinfo{year}{2005}): \emph{\bibinfo{title}{Tool Support for Checking
  Railway Interlocking Designs}}.
\newblock In: {\sl \bibinfo{booktitle}{Proceeding of the 10th Australian
  Workshop on Safety Related Programmable Systems (SCS'05)}},
  \bibinfo{publisher}{Australian Computer Science Communications}, pp.
  \bibinfo{pages}{101--107}.

\bibitemdeclare{inproceedings}{KWNR2003:ACSC}
\bibitem{KWNR2003:ACSC}
\bibinfo{author}{K.~\surnamestart Winter\surnameend} \&
  \bibinfo{author}{N.~\surnamestart Robinson\surnameend}
  (\bibinfo{year}{2003}): \emph{\bibinfo{title}{Modelling Large Railway
  Interlockings and Model Checking Small Ones}}.
\newblock In: {\sl \bibinfo{booktitle}{Proceeding of the Australian Cumputer
  Science Conference (ACSC 2003)}}.

\end{thebibliography}
\end{document}